\documentclass[aps,prd,twocolumn,groupedaddress,nofootinbib]{revtex4-1}
\usepackage{graphicx}

\begin{document}


\title{Origin of probabilities and their application to the multiverse}


\author{Andreas Albrecht}
\affiliation{University of California at Davis;
Department of Physics\\
One Shields Avenue;
Davis, CA 95616\\
}
\author{Daniel Phillips}
\affiliation{University of California at Davis;
Department of Physics\\
One Shields Avenue;
Davis, CA 95616\\
}



\begin{abstract}
We argue using simple models that all successful practical uses of probabilities
originate in quantum fluctuations in the microscopic physical world
around us, often propagated to macroscopic scales.  Thus we claim
there is no physically verified fully classical theory of
probability.  We comment on the general implications of this view, and
specifically question the application of purely classical probabilities
to cosmology in cases where key questions are known to have no quantum
answer. We argue that the ideas developed here may offer a way out of
the notorious measure problems of eternal inflation. 

\end{abstract}


\maketitle

\section{Introduction}
\label{Introduction}
We use the concept of probability extensively in science, and very
broadly in everyday life.  Many probabilistic tools used to ``quantify our ignorance'' seem 
intuitive even to non-scientists.  For example, if we consider
the value of one bit which we know nothing about,
we are inclined to assign probabilities to each value.  Furthermore,
it seems natural to give it a ``$50$-$50$'' chance of being $0$ or
$1$.  This everyday 
intuition is often believed to have deep theoretical justification based in ``classical
probability theory'' (developed in famous works such as~\cite{Laplace:1774zz}).  

Here we argue that the success of such
intuition is fundamentally rooted in specific physical properties of the
world around us.  In our view the things we call ``classical
probabilities'' can be seen as originating in the quantum probabilities that govern the microscopic
world, suitably propagated by physical processes so as to be
relevant on classical scales. From this perspective the validity of
assigning equal probabilities to the two states of an unknown bit
can be quantified by understanding the particular physical processes
that connect quantum fluctuations in the microscopic world to that
particular bit. The fact that we have simple beliefs about how to
assign probabilities that do not directly refer to complicated
processes of physical propagation is simply a
reflection of the intuition we have built up by living in a world
where these processes behave in a particular
way. Our position has implications for how we use probabilities in general,
but here we emphasize 
applications to cosmology which originally motivated our interest in
this topic. Specifically, we question a number of applications of
probabilities to cosmology that are popular today.

Many physicists view
classical physics as something that emerges from a fundamentally
quantum world under the right conditions (for example in systems
large enough to have negligible quantum fluctuations and with suitable
decohering behavior) without the need for new fundamental physics
outside of the quantum theory\footnote{We personally take this ``fundamentally
  quantum'' view but our arguments go
  through for some (but not all) other interpretations of
  quantum mechanics}. Taking that point of view does not make the
claims in this paper trivial ones.  Yes, in that picture ``all physics is
fundamentally quantum'', but here we focus specifically on the origin of
randomness.  Consider a classical computer well engineered to prevent
quantum fluctuations of its constituent particles from
affecting the classical steps of the computation.  One could
model a fluctuating classical system on such a computer (e.g.
a gas of perfect classical billiards), but the fluctuations in such an
idealized classical gas would indeed be classical ones.  The appearance
of a given fluctuation would reflect information already encoded in
classical features of the initial state of the computation and
would {\em not} come from quantum fluctuations of the particles
making up the physical computer. 
We argue that the real physical world does not contain 
such perfectly isolated 
classical systems and that quantum uncertainty, not ignorance of
classical information dominates probabilistic behavior we
observe. (For the computer example just given, the quantum
uncertainties will enter when setting up the 
initial state.)

In Bayesian language, the probability of a theory $T$ being true given
a dataset $D$ is computed by combining the probability of $D$ given
$T$ (``$P(D|T)$'') with the ``prior probability'' ($P(T)$) assigned to
$T$.  Often $P(T)$ will include other data combined in a
similar way. Inputting new data over time produces a 
list of updated probabilities. The start of such a list always
requires a ``model uncertainty'' (MU) prior that provides a personal
statement about which model(s) you prefer. Expressions for $P(D|T)$ can be tested by
statistical analysis of data and good scientists (discussing well
designed experiments) should agree on how
to compute $P(D|T)$.  The MU prior is a personal choice which is not
built from a scientifically  
rigorous process. The quantity $P(D|T)$ describes randomness in
physical systems, whereas MU priors represent states of mind of
individual scientists.   This paper only treats $P(D|T)$ probabilities, not
MU priors.  A further indication of the deep differences between
$P(D|T)$ and MU priors is that the goal of science is to produced
sufficiently high quality data (and sufficient consensus about the
theories) that which MU priors the community are willing to take is of no consequence to the
result. On the other hand, results will always depend strongly on at least some parts of
$P(D|T)$.

\section{The Page Problem}
\label{Page}
We outline the relevance of this question to cosmology using a simple
toy model. It is commonplace in cosmology to
contemplate a ``multiverse'' (e.g. in the context of ``eternal
inflation''~\cite{Guth:2007ng}) in which many equivalent copies of a given observer
 appear in the theory. 
As pointed
out by Page~\cite{Page:2009qe}, even if one knew the full wavefunction for
such a theory it would be impossible to make predictions about
future observations using probabilities derived from that
wavefunction.  The problem arises because multiverse theories
are expected to contain many copies of the observer (sometimes said to
be in different
``pocket universes'') that are identical in terms of
all current data, but which differ in details of their environments
that affect outcomes of future
experiments (e.g. experiments measuring neutrino masses or
cosmological perturbations).  In these theories it is impossible
to construct appropriate projection operators to describe measurements
where one does not know which part of the Hilbert space (i.e. which copy of
us and our world) is being measured. Thus, the outcomes of future
measurements are ill-posed quantum questions which cannot be answered
within the theory.

To illustrate this problem consider a
system comprised of two two-state subsystems called ``$A$'' and ``$B$''.
The whole system is spanned by the four basis states constructed as
products of basis states of the two subsystems: $\left\{ {{\left| 1
    \right\rangle }^{A}}{{\left| 1 \right\rangle }^{B}},{{\left| 1
    \right\rangle }^{A}}{{\left| 2 \right\rangle }^{B}},{{\left| 2
    \right\rangle }^{A}}{{\left| 1 \right\rangle }^{B}},{{\left| 2
    \right\rangle }^{A}}{{\left| 2 \right\rangle }^{B}} \right\}$.
For the whole system in state $\left| \psi  \right\rangle$, the
probability assigned to measurement outcome ``$i$'' can be
expressed as $\left\langle  \psi  \right|\hat{P}_i\left| \psi
\right\rangle $ for a suitably chosen projection operator
$\hat{P}_i$.  One can readily construct projection
operators corresponding to measuring system ``$A$'' in the ``$1$'' state
(regardless of the state of the ``$B$'' subsystem):
\begin{equation}
\hat{P}_{1}^{A}\equiv \left( {{\left| 1 \right\rangle
  }^{A}}{{\left| 1 \right\rangle }^{B}}{}^{B}\left\langle  1
\right|{}^{A}\left\langle  1 \right| \right)+\left( {{\left| 1
    \right\rangle }^{A}}{{\left| 2 \right\rangle
  }^{B}}{}^{B}\left\langle  2 \right|{}^{A}\left\langle  1 \right|
\right).
\end{equation} A similar operator $\hat{P}_{1}^{B}$ represents
measurements of only subsystem ``$B$''.  Operators such as $\hat{P}_{12} \equiv  {{\left| 1 \right\rangle
  }^{A}}{{\left| 2 \right\rangle }^{B}}{}^{B}\left\langle  2
\right|{}^{A}\left\langle  1 \right|$ represent measurements of {\em
  both} subsystems.

The problem arises because there is no projection operator that
gives the probability of outcome~``$1$'' when the subsystem to be
measured (``$A$'' or ``$B$'') is undetermined.  That is an ill-posed
question in the quantum theory.  Page emphasizes that this
kind of question apparently needs to be addressed in order to make
predictions in the multiverse, where our lack of knowledge about which
pocket universe we occupy corresponds to ``$A$'' vs. ``$B$'' not being
determined in the toy model. Such ill-posed quantum questions exist
in laboratory situations as well.  We tend not to be concerned about
these questions however, since there are also plenty of well-posed problems
on which to focus our attention.  Also, in the laboratory one might
resolve the problem by adding a measurable ``label'' to the setup that
does identify ``$A$'' vs. ``$B$''.  But such a resolution is believed
not to be 
possible in many cosmological cases. 

A natural response to this issue is to appeal to classical
ideas about probabilities to ``fill in the gap''.  In
particular, if one could assign classical probabilities $p_A$ and
$p_B$ 
for the measurement to be made on the respective subsystems,
then one could answer the question posed above (the probability of the
outcome ``$1$'' with the 
subsystem to be measured undetermined) by giving:
\begin{equation}
{{p}_{1}}={{p}_{A}}\left\langle  \psi  \right|\hat{P}_{1}^{A}\left|
\psi  \right\rangle +{{p}_{B}}\left\langle  \psi
\right|\hat{P}_{1}^{B}\left| \psi  \right\rangle.
\label{p1}
\end{equation}
Note that the values of  $p_A$ and $p_B$ are {\em not} determined from
$\left| \psi  \right\rangle$, and instead provide additional
information introduced to write
Eqn. \ref{p1}. Although $p_1$ can be written as the expectation value
of
${{\hat{P}}_{1}}={{p}_{A}}\hat{P}_{1}^{A}+{{p}_{B}}\hat{P}_{1}^{B}$,
the operator $\hat{P}_1$ is not a projection operator
($\hat{P}_1\hat{P}_1\neq\hat{P}_1$), confirming that $p_1$ does not give
probabilities of fully quantum origin.

Authors who apply expressions like Eqn. \ref{p1} to
cosmology~\cite{Srednicki:2009vb,*Page:2012gh} do not claim this gives  
a quantum probability.  Instead they appeal to classical
notions of probability along 
the lines we have discussed at the start of this paper. Surely one 
successfully introduces classical probabilities such as $p_A$ and
$p_B$ all the time in everyday situations to quantify our ignorance,
so why should the same approach not be used in the cosmological case?

Our view is that the two cases are completely different.  We
believe that in every situation where we use ``classical''
probabilities successfully to describe physical randomness these probabilities could in principle be
derived from a wavefunction describing the full 
physical situation.  In this context classical probabilities are just ways to
estimate quantum probabilities when calculating
them directly is inconvenient. Our
extensive experience using classical probabilities in this way (really
quantifying our {\em quantum} ignorance) cannot 
be used to justify the use of classical 
probabilities in situations where quantum probabilities have been
clearly shown to be ill-defined and uncomputable. Translating the
formal framework from one situation to the other is not an extrapolation
but the creation of a brand new conceptual framework that needs
to be justified on its own\footnote{Cooperman~\cite{Cooperman:2010zc}
  has explored the interpretation of these matters in the context of
  the Positive Operator Valued Measure (POVM) formalism. In our view
  this does not really resolve the problem, since one has to introduce
  new probabilities equivalent to $p_A$ and $p_B$ in an equally ad hoc
  way.  We definitely do agree with the connections he draws to the
  standard treatment of identical particles, which we find quite intriguing.}.

We are only challenging the ad hoc introduction
of classical probabilities such as $p_A$ and $p_B$. We are not criticizing
the use of standard ideas from probability theory to manipulate and
interpret probabilities that have a physical origin.
Of course we never know the wavefunction completely (and thus often
write states as density matrices). Our claim is that probabilities are only
proven and reliable tools if they have clear values determined from the quantum
state, despite our uncertainties about it.  


\section{Billiards}
\label{Billiards}
We next use simple calculations to argue that it is realistic
to expect all probabilities we normally use to have a quantum origin.
Consider a gas of idealized billiards with radius $r$, mean free path
$l$,average speed ${\bar v}$ and mass $m$. If two of these billiards
approach each other with impact parameter $b$, the uncertainties in the
transverse momentum ($\delta {{p}_{\bot }}$) and position ($\delta
{{x}_{\bot }}$) contribute to an uncertainty in the impact parameter given by:
\begin{equation}
   \Delta b  
 =\delta {{x}_{\bot }}+\frac{\delta {{p}_{\bot}}}{m}\Delta t
 =\sqrt{2}\left( a+\frac{\hbar }{2a}\frac{l}{m\bar{v}} \right) 
\label{Eqn:Deltab}
\end{equation}
where the second equality is achieved using $\Delta t = l/{\bar
  v}$ and assuming a minimum uncertainty wavepacket of width $a$ in
each transverse direction. The value of $\Delta b$ is
minimized by $a=\sqrt{\hbar l /(2m{\bar v})} \equiv \sqrt{l
  \lambdabar_{dB}/2}$. We will show that $\Delta b$ is
significant even when minimized.

The local nature of subsequent collisions creates a distribution of entangled
localized states reflecting the range of possible collision points
implied by $\Delta b$. We estimate the width of this distribution as
it fans out toward the next collision by classically propagating
collisions that occur at either side of the range $\Delta
b$. (Neglecting additional quantum effects increases the
robustness of our argument.)
The geometry of the collision amplifies uncertainties in a manner
familiar from many chaotic 
processes~\cite{Birk27a,Zurek:1994wd}.  The quantity 
$\Delta b_{n} =\Delta b( 1+(2l)/r)^{n}$
gives the uncertainty in $b$ after $n$ collisions.

Setting $\Delta b_n=r$ and solving for $n$ determines $n_Q$,
the number of collisions after which the quantum spread is so large that
there is significant quantum uncertainty as to which billiard takes
part in the next collision:
\begin{equation}
{{n}_{Q}}=-\frac{\log \left( \frac{\Delta b}{r} \right)}{\log \left(
  1+\frac{2l}{r} \right)}.
\label{Eqn:nQ}
\end{equation}
 For Table
\ref{Table} we evaluated Eqn. \ref{Eqn:nQ} with different input
parameters chosen to represent various physical
situations.\footnote{Raymond~\cite{Raymond:1967aa} presents similar
  result, applied only to actual billiards.  He also makes some
  general points about the implications of his result that overlap
  with some of the points we are making here.}
\begin{table*}[htbp]

    \begin{tabular}{l|r|r|r|r|r|r|r|}

      & \multicolumn{1}{|c|}{$r$ {\it(m)}} 
      & \multicolumn{1}{|c|}{$l$ {\it(m)}} 
      & \multicolumn{1}{|c|}{$m$ {\it (kg)}} 
      & \multicolumn{1}{|c|}{${\bar v}$ {\it (m/s)}} 
      & \multicolumn{1}{|c|}{$\lambdabar_{dB}$ {\it (m)}}
      & \multicolumn{1}{|c|}{$\Delta b$ {\it (m)}}
      & \multicolumn{1}{|c}{$n_Q$} \\ \hline

    Nitrogen at STP (Air) & $1.6 \times 10^{-10}$ & $3.4\times
    10^{-07}$ & $4.7\times 10^{-26}$ & $360$
    &$ 6.2 \times 10^{-12}$ & $2.9\times 10^{-9}$ & $-0.3$ \\ \hline
    Water at body temp & $3.0\times 10^{-10}$ & $5.4 \times 10^{-10}$
    & $3.0\times 10^{-26}$ & $460$ &
    $7.6\times 10^{-12}$ & $1.3 \times 10^{-10}$ & $0.6$ \\ \hline
    Billiards game& $0.029$ & $1$     & $0.16$  & $1$ & $6.6 \times
    10^{-34}$ & $5.1 \times 10^{-17}$ & $8$
    \\ \hline
    Bumper car ride & $1$     & $2$     & $150$   & $0.5$ & $1.4\times
    10^{-36}$ & $3.4\times 10^{-18}$& $25$
    \\ \hline

    \end{tabular}%
  \caption{The number of collisions, ($n_Q$ from  Eqn. \ref{Eqn:nQ})
    before quantum uncertainty dominates, evaluated for physical
    systems modeled as a ``gas'' of billiards with
    different properties. Values  $n_Q < 1$
    indicate that quantum fluctuations are so dominant that
    Eqn. \ref{Eqn:nQ} breaks down.  All randomness in
    these quantum dominated systems is fundamentally quantum in nature.  \label{Table}}
\end{table*}%

Table \ref{Table} shows that water and air are so dominated by quantum fluctuations
that $n_q < 1 $, indicating the breakdown of Eqn. \ref{Eqn:nQ}, but
all the more strongly supporting our view that {\em all} randomness in
these systems is fundamentally quantum. This result strongly indicates
that if one were able to fully
model the molecules in these macroscopic systems one would find that the
intrinsic quantum uncertainties of the molecules, amplified by
processes of the sort we just described,  would be fully
sufficient to account for all the fluctuations.
One would not be required to ``quantify our ignorance'' using 
classical probability arguments to fully understand the system.  For
example, the Boltzmann distribution for one of these systems in a
thermal state should really be derivable as a feature dynamically
achieved by the wavefunction without appeal to formal arguments about
equipartition etc. 

This argument that the randomness in collections of molecules in the world
around us has a fully quantum origin lies at the core of our case.  We
expect that all practical applications of probabilities can be traced
to this intrinsic randomness in the physical world.   As an 
illustration, we next trace the randomness of a coin
flip to Brownian motion of polypeptides in the human nervous system. 

\section{Coin Flip}
\label{Coin}

Randomness in a coin flip comes from a lack of correlation between the
starting and ending coin positions.   The
signal triggering the flip travels along
human neurons which have an intrinsic
temporal uncertainty  of $\delta t_n \approx 1ms$~\cite{Faisal2008}. 
It has been argued that fluctuations in the number of open neuron ion channels can account for the
observed values of $\delta t_n$~\cite{Faisal2008}.   These molecular fluctuations are due to random Brownian motion
of polypeptides in their surrounding fluid.  Based on our assessment that the
probabilities for fluctuations in water are
fundamentally quantum, we argue that the value of $\delta t_n$
realized in a given situation is also fundamentally quantum. Quantum
fluctuations in the water drive the motion 
of the polypeptides, resulting in different numbers of ion
channels being open or closed at a given moment in each instance
realized from the many quantum possibilities. 

Consider a coin flipped and caught at about the same height,
by a hand moving at speed $v_h$ in the direction of 
the toss and with a flip
imparting an additional speed $v_f$ to the coin. A neurological
uncertainty in the time of the flip, $\delta t_n$, results in a
change in flight time $\delta t_f = \delta t_n \times v_h/(v_h+v_f)$.  A
similar catch time uncertainty gives a total flight time uncertainty
$\delta t_t = \sqrt{2} \delta t_f$.  A coin flipped upward by an
impact at its edge has a rotation frequency 
$f=4v_f/(\pi d)$ where $d$ is the coin diameter. 
The uncertainty in the
number of spins is $\delta N = f \delta
t_t$. Using $v_h=v_f=5m/s$ and $d=0.01m$ (and $\delta t_n = 1ms$) gives $\delta N = 0.5$, 
enough to make the outcome of the coin toss completely dependent on
the time uncertainty in the neurological signal which we 
have argued is fully quantum.

No doubt we have neglected significant factors in 
modeling the coin flip. 
The point here
is that even with all our simplifications, we 
have a plausibility argument that the outcome of a coin flip is truly
a quantum measurement (really, a Schr\"{o}dinger cat) and that the
$50$--$50$ outcome of a coin toss may in principle be derived from
the quantum physics of a realistic coin toss with no reference to
classical notions of how we must ``quantify our ignorance''.
Estimates such as this one illustrate how the quantum nature of
fluctuations in the gasses and fluids around us 
can lead to a fundamental quantum basis for probabilities we
care about in the macroscopic world. 

\section{Digits of $\pi$}
\label{Digits}

The view that all practical applications of probabilities are
based on physical quantum probabilities 
seems a challenging proposition to
verify.  As we have illustrated with the coin flip, the path from
microscopic quantum fluctuations  
to macroscopic phenomena is complicated 
to track. 
And there are
endless cases to check (rolling dice, 
choosing a random card etc.), most 
also too complicated to work through conclusively.  So arguing 
our position on a case-by-case basis is certainly an impractical
task. 

On the other hand, our
ideas are very easy to falsify. All one needs is one illustration of a
case where classical notions of probability are useful in a physical
system that is fully isolated from the quantum fluctuations. Once the
practical value of purely classical 
probabilities is established there is no reason 
it should not be applicable to other situations. One idea for such a counterexample was proposed
by Carroll.\footnote{S. Carroll at the {\em PCTS
  workshop on inflation} (Jan 2011).} One could place bets on, say,
the value of the millionth digit of $\pi$.  Since the digits of
$\pi$ are believed to be random~\cite{Bailey:1997xx} one should be
able to use this apparently purely classical notion to win bets.
While on the face of it this appears to be an 
ideal counterexample, further scrutiny reveals an essential quantum role. 

Let's phrase this problem more systematically:  One expects that if you
finds someone who thinks the digits of $\pi$ are not randomly
distributed, you can make money betting against them. Or
equivalently, the expected payout $P_\pi$ is zero if betting with someone who
{\em does} think the digits are random. A simple formula for such a
payout is given by
\begin{equation}
 P_\pi   =  \lim_{N_{tot} \to \infty} {1 \over N_{tot}} \sum_{\{i\}} \left( N_{\pi}^i - 4.5
\right) = 0 
\label{BetOnPi}
\end{equation}
where $\{i\}$ is the ensemble (of size $N_{tot}$) of the digits chosen and $N_\pi^i$ is the actual
value of the $i$th digit of $\pi$. The result depends entirely on
the choice of ensemble.  With enough knowledge of $\pi $ one can
come up with ensembles that give any answer you like (for example that
only ever select the digit ``$1$''), despite all the randomness
``intrinsic'' to $\pi$ (and in fact {\em because} 
the properties of $\pi$ are classical and knowable).  Thus
we argue that the outcomes of such bets are all about the ensemble
selected, and the choice of the ensemble is the only source of
randomness in the entire activity. 

The reason the initial idea of betting on $\pi$ is so
compelling is that no one ever thinks an ensemble will be chosen with
attention to the actual values of the digits of $\pi$. One can see
how quantum mechanics comes in by scrutinizing the process of coming
up with ensembles. 
It could be through the human neurons used in selecting a classical
random number seed\footnote{Similarly, the involvement of neurons etc. with the
initial setup prevents the classical computer example in
Sect. \ref{Introduction} from being a counterexample.}, or through something
more systematic like a roulette wheel.  Again this falls in the
category where one counterexample could ruin the argument, but so far
we have not found one.    The bet really is
about the lack of correlation between the digit selection and the
digit value and we argue it is quantum processes such as those discussed
here that are being counted on to create the lack of correlation that
is crucial to the fairness of the bet.  


Our analysis depends crucially on seemingly
``accidental'' levels of quantum noise in the physical world. Our
point is that accidental or not, we count on
this quantum noise to produce the uncorrelated
microscopic states that lie at the heart of our understanding of
randomness and probabilities in the world around us. Extending this
understanding to domains where quantum noise cannot play this role is
not at all straightforward.  Discussions 
of the non-random behaviors of classical random number generators
(such as in~\protect\cite{Press:1992zz}) underscore the difficulty
of even imagining a classical source of randomness with the necessary
lack of correlations. 

\section{Toward a solution of cosmic measure problems}
\label{Applications}
So far we have used our ideas about probability to critique
the introduction of purely classical probabilities into cosmological
theories, which is an approach advocated by others~\cite{Srednicki:2009vb,*Page:2012gh}. In this section we
use the ideas introduced here to work out our own 
approach to probabilities in the multiverse.   
We embrace the idea advocated above, 
that fundamentally classical probabilities have no place in
cosmological theories, and declare that questions that seem to
require classical probabilities for answers simply are not answered in
that theory. We are basically advocating a more strict discipline
about which questions are actually addressed by a given theory.\footnote{Although here we focus on cosmology, it appears that
  our approach is relevant to other areas where there is confusion
  about about how to assign probabilities, such as the ``sleeping beauty problem''\cite{Elga01042000}.}
Then one can ask if there are multiverse theories with sufficient
predictive power to remain viable after this discipline is
imposed. Our first assessment of this question suggests that imposing 
this discipline may reduce or completely eliminate the notorious
measure problems of eternal inflation and the multiverse. 

One challenge one faces when exploring this matter is the fact that most
discussions of eternal inflation and the multiverse are approached in
a semiclassical manner (for example assuming well-defined
classical spatial slices of infinite extent).  A more careful attempt
to identify the full quantum nature of the picture may point to
additional ways proper quantum probabilities are assigned. We will not
try to address that aspect of the question here, and really just take a
first look at the impact of hewing to our proposed probability
discipline. 

A general point immediately becomes clear:  We are used to
linking counting with probabilities, but such connections are not 
always direct or relevant.  Counting up the heads and tails in a long string of
coin flips {\em is} connected with proper quantum
probabilities. Starting with our results of Sect.~\ref{Coin} one can
see that a specific quantum probability is assigned to each different
possible heads/tails count, and thus counting can be tied in to
well-defined quantum probabilities for that system.  However, the fact that one
cosmology may have $3$ pocket universes of type $A$, while another may
have $10^{100}$ does not make a difference, because as we discussed in
Sect.~\ref{Page}, no quantum probabilities can be constructed to
determine which among different (equivalent so far) observers you
might be.  While these numbers (by analogy with the flips of multiple
coins) may be linked to global properties of the state, they cannot by
used to determine which among equivalent patches a given observer occupies.  

The insight that counting of observers in itself is insufficient to
lead to proper probabilities leads to some interesting conclusions. 
One is immediately drawn to the question of ``volume factors'' that give large volume regions more weight than
small ones.  To the extent that volume factors are only a stand-in for
counting observers we regard such counting as meaningless because it 
cannot be related to true quantum probabilities. 

This insight also relates to the ``young universe'' or ``end of time''
problem~\cite{Bousso:2010yn,Guth:2011ie}, which can be sketched as follows:
If one regulates the 
cosmology with a time cutoff, inflation guarantees that most pocket
universes will be produced close to the cutoff. Then the
time cutoff shows up at early times (relative to their time of
production which is under strong pressure to happen late) for most
pocket universes.  This problem persists even as one pushes the time
cutoff out to infinity.  But there is no evidence 
that this counting has anything to do with probabilities
predicted by the theory which are relevant to an observer.  There is
no sign that such theories are able to assign a true quantum
probability to the time when a particular observer's pocket 
universe was produced.  One is simply looking at different pocket
universes, and which one we occupy is not determined by the theory.

Our position appears to offer significant implications for the Boltzmann Brain
problem~\cite{Albrecht:2014eaa,Albrecht:2004ke,Page:2006ys}.  For our purposes here,
this problem is simply the case where pathological
observers, called Boltzmann Brains or BB's, vastly outnumber realistic
ones.  (The pathology of the BB's is that they match all the data we
have so far, but the next moment experience catastrophic breakdown
of physicality, experiencing a rapid heat death.)
Again, we claim here that counting numbers of BB's vs realistic
observers cannot be related to quantum probabilities predicting which an observer
is more likely to experience.
Thus, as long as there is at least one realistic pocket universe,
there will be no BB problem, no matter how many BB's are produced in
the theory. 

Now let us look at this matter from a slightly different point of
view.  The real problem arises when one does not know which part of
the Hilbert space one is about to measure.  However, if one just takes
one piece of the Hilbert space in an eternally inflating universe,
that patch alone will have probabilities of tunneling into pocket
universe $A$ or $B$, and perhaps many other outcomes as well. If one
simply traces out the rest of the Hilbert space, one will have a
density matrix for what is going on in that patch.  With that one {\em
  can} take expectation values of operators, without introducing
classical probabilities to determine which pocket you are in. To the
extent that the BB problem can be phrased in this way (in terms of a
quantum branching into BB's vs realistic cosmologies in a given patch),
we expect the BB problem will remain if realistic cosmologies are
sufficiently suppressed\footnote{In \cite{Albrecht:2014eaa} one of us (AA) treats
  BB's in the traditional counting language in toy models.  However, we
expect that with a bit more realism the kind of quantum chaos discussed in
this paper would allow those BB discussions to go over nicely into
the (more legitimate) quantum branching form described here, without changing the
conclusions in~\cite{Albrecht:2014eaa}.}. And if
all patches are the same (as may well be the case for highly symmetric
theories such as eternal inflation) then it does not really matter
what patch you are in.  The answer will still be the same. 

While we
have yet to offer a rigorous demonstration, this set of ideas seem
promising to us as a way out of the measure problems in cosmology.  A
more formal way to describe this picture is that if one does consider
a theory with multiple possible locations for the observer, one would
be obliged to give a ``prior'' on which location we occupy.  These
priors would look very much the same as the classical probabilities
that show up for example in Eqn.~\ref{p1}. However, by viewing these
probabilities as priors,
our agenda would be to reach a point where their values do not matter
to our answers\footnote{Note that while formally these priors look the
  same as the classical probabilities discussed
  in~\cite{Srednicki:2009vb,*Page:2012gh}, those authors emphasize
  cases where results {\em do} depend in a fundamental way on the values chosen
  for the classical probabilities.  So they
  are not really treating their classical probabilities
  as prior probabilities, the values of which should ultimately not be
  important.}. It would 
appear that for sufficiently 
symmetric theories, independence from these priors would be easy to
achieve.  Also, if certain obserables are sufficiently correlated, the
measurement of one (which itself did not have a prediction for the
outcome due to dependence on priors) could then lead to predictions
for the other observable. Both of these pictures outlined here could
lead to a substantial level of predictive power, despite the
restrictions imposed by our probability discipline. 

\section{Conclusions}
\label{Conclusions}

In summary, we have argued that all successful applications of
probability to describe nature can be traced to quantum origins.
Because of this, there has not been any systematic validation
of purely classical probabilities, even though we appear to 
use them all the time. These matters are of particular importance in 
multiverse theories where truly classical probabilities are used
to address critical
questions not addressed by the quantum theory.   
Such applications of classical probabilities need to be built
systematically on separate foundations and not be thought of as
extensions of already proven ideas. 
We have yet to see purely classical probabilities motivated and
validated in a compelling way, and thus are skeptical of
multiverse theories that depend on classical probabilities for their
predictive power. Fundamentally finite cosmologies~\cite{Banks:2003pt,*Albrecht:2011yg} that
do not have duplicate observers do not require classical
probabilities. These seem to be a more promising path. 

We are not the only ones who regard quantum
probabilities as most fundamental
(e.g.~\cite{Deutsch:1999gs,*Wallace:2010aa,*Zurek:2011zz,*Bousso:2011up}), but there
are also opposing views\footnote{There are also some papers 
  where the degree overlap is not so clear.  Vilenkin appears to focus on
  quantum probabilities in~\cite{Vilenkin:2013loa}, but then also
  seems to embrace a fundamentally classical picture similar to that
  advocated in~\cite{Aguirre:2010rw}. Some aspects
  of~\cite{Nomura:2011rb} also seem to overlap, although other things
  (such as the emphasis on holography) seem very different, so it is
  hard to tell the overall degree of agreement}.  In addition to the
case already discussed where classical probabilities are introduced in
multiverse theories to enhance predictive 
power (such
as in~\cite{Srednicki:2009vb,Page:2012gh}),
some theories insert classical
ideas for other reasons, often in hopes of allaying
interpretational concerns
(e.g. ~\cite{'tHooft:2010zz,Weinberg:2011jg,Aguirre:2010rw}).
The arguments presented here make us generally 
doubtful of such classical formulations, since our analysis reinforces the
fundamental role of quantum theory in our overall understanding
of probabilities. Perhaps some of these alternate theories integrate the
classical ideas sufficiently tightly with the quantum piece that the
everyday tests we have discussed could just as well be regarded as
tests of the classical ideas in the alternate theory.  However, such 
logic seems overly complex to us, and we prefer the simpler
interpretation that the strong connection between all our
experiences with probabilities and the quantum world means the quantum
theory really is the defining physical theory of probabilities.  We
have offered suggestions that sticking only to quantum probabilities
to make predictions in the multiverse may not be all that debilitating to
the predictive power of multiverse theories and may actually offer a
solution to the notorious measure problems of eternal inflation.





\acknowledgments
We thank E.~Anderes, A.~Arrasmith, S.~Carroll, J.~Crutchfield, D.~Deutsch, 
B.~Freivogel, A.~Guth, 
J.~Hartle, T.~Hertog, T.~Kibble, L.~Knox, Z. Maretic, D.~Martin, J.~Morgan, J.~Preskill,
R.~Singh, A.~Scacco, M.~Sredniki, A.~Vilenkin and W.~Zurek for
helpful conversations.  We were supported in part by
DOE Grant DE-FG03-91ER40674.

\bibliography{AA}

\end{document}